\documentclass[a4paper]{jpconf}
\usepackage{graphicx}
\begin{document}
\title{Inelastic neutron scattering studies of the quantum frustrated magnet clinoatacamite,  $\gamma$-Cu$_2$(OD)$_3$Cl, a proposed valence bond solid (VBS)}

\author{A. S. Wills$^{1,2}$, T. G. Perring$^3$, S. Raymond$^4$, B. F{\aa}k$^4$, J.-Y. Henry$^{4}$ and M. Telling$^3$}

\address{$^1$Department of Chemistry, University College London,
20 Gordon Street, London, WC1H 0AJ, UK}
\address{$^2$The London Centre for Nanotechnolgy, 17-19 Gordon Street, London WC1H 0AH, UK}
\address{$^3$% 
ISIS Facility, Rutherford Appleton Laboratory, Rutherford Laboratory, Chilton, Didcot, OX11 0QX, UK}
\address{$^4$% 
Commissariat \`a l'Energie Atomique, INAC,SPSMS, 38054 Grenoble, France}

\ead{a.s.wills@ucl.ac.uk}

\begin{abstract}

The frustrated magnet clinoatacamite, $\gamma$-Cu$_2$(OH)$_3$Cl, is attracting a lot of interest after suggestions that  at low temperature it forms an exotic quantum state termed a Valence Bond Solid (VBS) made from dimerised Cu$^{2+}$ ($S=\frac{1}{2}$) spins.\cite{Lee_clinoatacamite} Key to the arguments surrounding this proposal were suggestions that the kagom\'e planes in the magnetic pyrochlore lattice of clinoatacamite are only weakly coupled, causing the system to behave as a quasi-2-dimensional magnet. This was reasoned from the near 95$^\circ$ angles made at the bridging oxygens that mediate exchange between the Cu ions that link the kagom\'e planes.

Recent work pointed out that this exchange model is inappropriate for  $\gamma$-Cu$_2$(OH)$_3$Cl, where the oxygen is present as a $\mu_3$-OH.\cite{Wills_JPC} Further, it used symmetry calculations and neutron powder diffraction to show that the low temperature magnetic structure ($T<6$~K) was canted and involved significant spin ordering on all the Cu$^{2+}$ spins, which is incompatible with the interpretation of simultaneous VBS and N\'eel ordering.  Correspondingly, clinoatacamite is best considered a distorted pyrochlore magnet. In this report we show detailed inelastic neutron scattering spectra and revisit the responses of this frustrated quantum magnet.

\end{abstract}

%\section{Introduction}

Frustration in magnetic systems is an underlying theme that links together many of the properties that are captivating modern condensed matter science. It creates degeneracies and destabilises conventional order, allowing exotic effects to occur, such as spin ices,\cite{spin_ice,Kag_spin_ice} colossal magneto-resistance,\cite{ZnCuCrSe} topological spin glasses\cite{CanJPhys} and the anomalous Hall effect.\cite{Hall} Despite the extensive theoretical studies made over the last two decades, little is known about the effects of frustration in experimental systems in the S=1/2 quantum limit due to the rarity of useful model materials. In particular, model systems based on the pyrochlore lattice: a network of corner-sharing tetrahedra that provides the framework upon which many frustrated 3-D magnets are based.\cite{Gd2Sn2O7,Er2Ti2O7}

%rewrite
Previous studies of clinoatacamite showed it to be frustrated with the strong exchange,  $\theta_{CW} \leq 180$~K leading to ordering transitions at the suppressed temperatures of $T_{C2}\sim6$~K\cite{Wills_ICM} and $T_{C1}\sim18$~K\cite{Japanese_PRB, Japanese_PRL}. The small amount of entropy released at both of these transitions is indicative of a ground state with extensive fluctuations.  A canted and weakly ferromagnetic spin structure has been observed below the sequential transitions at 6.1 and 5.8~K.\cite{Wills_JPC} The details of the 18~K transition are less clear. Recently, Lee {\it et al}. proposed that the elastic and inelastic neutron scattering spectra are compatible with the formation of a valence bond solid (VBS) at 18~K that survives the formation of collinear magnetic order at $\sim6$~K.\cite{Lee_clinoatacamite} This is hard to reconcile with the ferromagnetism and canting observed in the low temperature magnetic structure\cite{Wills_JPC} and the static components observed in the MuSR spectra.\cite{Japanese_PRL}   The proposal of the VBS  state arose from 
arguments, now shown to be unfounded,\cite{Wills_JPC}  that  the geometry of the superexchange leads to an effective decoupling of the 3-D distorted pyrochlore into separated kagom\'e planes, reminiscent of the responses seen in Li$_2$Mn$_2$O$_4$ and the spinel series Li$_x$Mn$_2$O$_4$.\cite{Li2Mn2O4,LixMn2O4}

%\section{Crystal structure of clinoatacamite}

The main details of the crystal structure of clinoatacamite were determined by Oswald and Guentier.\cite{Oswald} A complete description, including the positions of the deuteriums, are given in a recent neutron powder diffraction study.\cite{Wills_JPC} The crystal structure of clinoatacamite is distorted away from the rhombohedral symmetry of the parent atacamite structure  due to the presence of a cooperative Jahn-Teller  effect, and is described in the monoclinic space group $P~1 ~ 2_1 ~n ~ 1$. Using the notation in \cite{Wills_JPC} the magnetic pyrochlore-like lattice is made up from chains of Cu1 ions that run parallel to the $b$-axis, and approximately perpendicular chains parallel to the $a$-axis made from alternating Cu2 and Cu3. A key feature for the magnetic exchange is the bridging $\mu_3$-hydroxide group. Defining the  Cu\--- $(\mu_3$-OH$)$\---Cu bridging angle as  $\phi$, angles in the range  $91<\phi<102^\circ$ are found in clinoatacamite which will mediate a range of competing ferromagnetic and antiferromagnetic interactions.

%\section{Method}

Inelastic neutron scattering experiments were performed using the thermal and cold energy time-of-flight (TOF) spectrometers MARI and OSIRIS respectively at the ISIS Facility, and IN22 of the ILL. 5g of fully deuterated $\gamma$-Cu$_2$(OD)$_3$Cl  powder were prepared following the method described in \cite{Wills_JPC} and held in a cylindrical aluminium can in an annular configuration for the TOF instruments and a cylindrical can for triple axis spectrometer IN22.

\begin{figure}
\includegraphics[width=35pc]{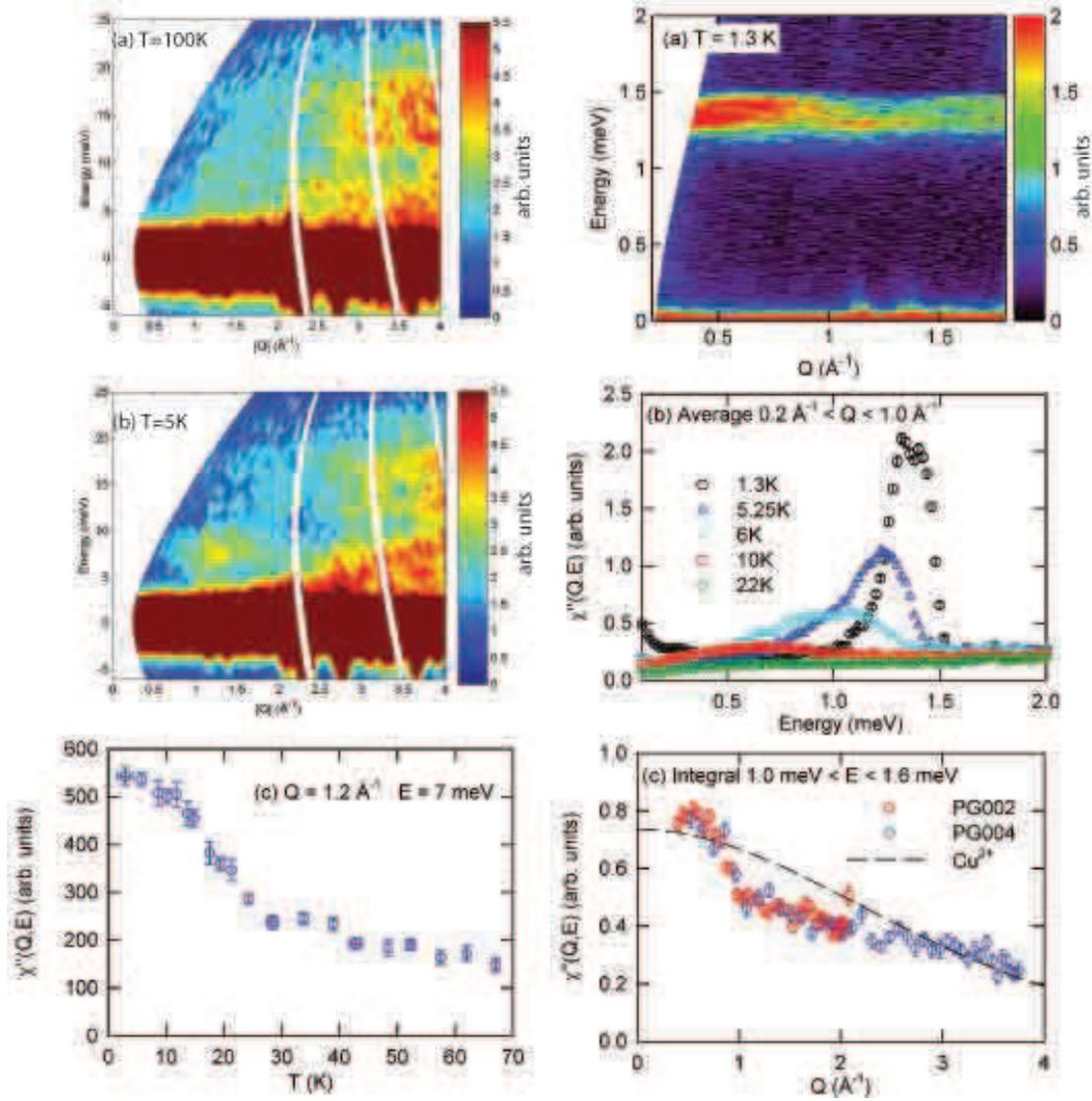}\hspace{2pc}%
\caption{\label{label}Inelastic neutron scattering spectra of clinoatacamite. (Left) The inelastic neutron scattering spectrum of $\gamma$-Cu$_2$(OD)$_3$Cl collected using the thermal time-of-flight spectrometer MARI shows the $Q$ and $\omega$-dependence of the diffuse excitation characteristic of the spin liquid phase centred at $Q\sim1$~\AA$^{-1}$ and $E\sim 7$~meV (data collected with incident neutron energy 40 meV) as a function of temperature (Figures a and b). Figure c shows the temperature dependence of $\chi''$ taken at $(Q,E)=(1.2$~\AA$^{-1},7$~meV). Right) The $Q$ and $E$ dependence of the low energy excitation spectrum measured using the cold neutron time-of-flight spectrometer OSIRIS. A nearly dispersionless band at $E=1.3$meV appears below $T_{C2}\sim6$~K. b) The integrated scattering between $0.2$\AA$^{-1} \leq Q \leq 1.0 $\AA$^{-1}$ show how this mode grows out from the elastic line upon cooling. c) This characteristic excitation of the low temperature phase $T<T_{C2}$ was isolated by subtraction of the data collected at 10K from that at 1.3K. A cut centred on $E = 1.3$ meV as a function of $Q$ does not follow the squared magnetic form factor dependence expected for free Cu$^{2+}$ ions (dashed line), indicating that the excitation involves correlated Cu$^{2+}$ spins. The data shown in a) and b) were collected using the 002 reflection of the pyrolytic graphite (PG) analysers and those shown in c) were collected using the 002 and 004 reflections as indicated. The analysed scattered neutron energies are 1.846 meV and 7.385 meV for the 002 and 004 reflections respectively.}
\end{figure}

%\section{Results and Discussion}

The first feature to note is the diffuse response in the time-of-flight inelastic spectra centred at $Q \sim 1$~\AA$^{-1}$, $E\sim7$~ meV (Figure 1). Its continued presence over the wide temperature range $1.6 < T <30$~K and the close agreement of the energy with the mean field result $<J>=3\theta_W/[2zS(S+1)] \sim -60$~K $\approx 5.2$~meV for a pyrochlore antiÂferromagnet with $z=6$, show it to be a fundamental response of this highly frustrated system and allows confirmation of the mean value of $J$. The reciprocal distance $Q\sim1$~\AA$^{-1}$ of this excitation corresponds to twice the mean nearest neighbour Cu-Cu distances, $2<r_{Cu-Cu}>~ 6.43$ \AA. Upon cooling to low temperatures only a small change is seen in the form of the scattering that corresponds to a broadening of the contribution towards low $Q$, signifying an increase in the spatial extension of these correlations in real space.

The observation of this diffuse excitation over such a wide range of temperature shows that it is a fundamental signature of the frustration in this quantum magnet. The insensitivity of both the energy responses and spatial extents of the excitation to temperature indicate that this low-temperature phase resembles the system above $T_{C1}\sim18$~K, and that the spins are therefore fluctuating rapidly in a spin liquid phase. The continued presence of this well-defined resonance below both $T_{C1}$ and $T_{C2}$ and the similarity between the responses at high and low temperatures demonstrates that the excitation is of fundamental clusters of spins that are robust to the formation of the magnetic orderings associated with both of these transitions. Its presence in all the zero-field magnetic phases of $\gamma$-Cu$_2$(OD)$_3$Cl shows that both magnetic phase transitions involve incomplete ordering. Such a juxtaposition of components with different orderings is a particular property of highly frustrated systems and occurs when subsets of the macroscopically degenerate ground state remain robust to the formation of some partial ordering. In \cite{Lee_clinoatacamite} this was attributed this diffuse scattering to singlet-triplet excitations of the spin dimers in the VBS state. Our data show that  the broadness of the response and the lack of structure prevents its form from being particularly characteristic of a single type of excitation and it is not possible to say anything conclusive about its origin without more detailed modelling.

Comparison of inelastic scattering data collected between 1.5 and 20~K using cold neutrons shows that while there is no change in the low energy $(Q, E)$ spectra at $T_{C1}\sim18$~ K, an excitation at 1.3~meV grows out of diffuse elastic scattering upon cooling below $T_{C2}\sim 6$ K. Our previous studies showed that this mode is a fundamental signature of the low temperature physics of clinoatacamite as its energy corresponds to the gap used to fit the thermally activated specific heat below 6~K.\cite{Wills_ICM} At first glance the excitation appears to be both non-dispersive and gapped by an energy of $\Delta\sim1.3$~meV in agreement with our earlier report. Close examination, however, reveals that there is a weak dispersion within the excitation band that gives it a half-width of $\sim0.1$~meV. The $Q$ dependence of the scattering integrated over the full band-width of the excitation shows that the excitations involve correlated spins as there is a deviation from the form factor dependence characteristic of a free Cu$^{2+}$ ion. 

The small dispersion given the large $<J>$ implies that the excitation cannot be described by standard spin wave theory in the limit of weak anisotropy and that such anisotropy needs to be taken into account in future investigations.  The well-defined nature and width of the band is remarkable given that the sample is a powder and raises the possibility that quantum effects are involved in the definition of a single sharp energy scale.  The weak intraband dispersion is indicative of additional further neighbour magnetic interactions and can be used directly to establish their energy scale as $\sim0.1$~meV. These findings, together with the deviation from the free ion form factor,  are generally at variance with the suggestion in \cite{Lee_clinoatacamite} relating the 1.3~meV response to single-ionic excitations of uncoupled spins.

In conclusion we show using inelastic neutron scattering that despite the substantial distortion to its structure, the low temperature physics of $\gamma$-Cu$_2$(OD)$_3$Cl  is dominated by the frustration associated with the underlying pyrochlore lattice. There are two distinct excitation types in the system: high energy short-ranged excitations (7 meV) characteristic of the spin liquid phase coexist at low temperature with well-defined low energy modes (1.3 ~meV).

\ack
ASW would like to thank the Marie-Curie project of the EU, the Royal Society and EPSRC (grant number EP/C534654) for financial support and Steve Bramwell, Peter Holdsworth and Philippe Mendels for discussions.

\end{document}